\documentclass[11pt,twoside]{article}
\usepackage{asp2014}

\usepackage{wrapfig}    

\aspSuppressVolSlug
\resetcounters

\begin{document}

\title{BoF: Interoperability of Users, Developers, and Managers}

\author{Jan~Noordam$^1$, Yan~Grange$^2$, and Peter~Teuben$^3$}
\affil{$^1$Madroon Community Consultants, De Marke 35, 7933 RA, Pesse, The Netherlands; \email{noordam@astron.nl}}
\affil{$^2$ASTRON, Netherlands Institute for Radio Astronomy, Oude Hoogeveensedijk 4, 7991 PD, Dwingeloo, The Netherlands}
\affil{$^3$University of Maryland, College Park, Maryland, USA}

\paperauthor{Jan~Noordam}{noordam@astron.nl}{ORCID}{Madroon Community Consultants}{}{Pesse}{The Netherlands}{7933 RA}{The Netherlands}
\paperauthor{Yan~Grange}{grange@astron.nl}{0000-0001-5125-9539}{ASTRON}{}{Dwingeloo}{}{7991PD}{The Netherlands}
\paperauthor{Peter~Teuben}{teuben@astro.umd.edu}{0000-0003-1774-3436}{University of Maryland}{Astronomy Department}{College Park}{MD}{20742}{USA}



 

\begin{figure}[ht]
    \centering
    \includegraphics[width=0.8\textwidth]{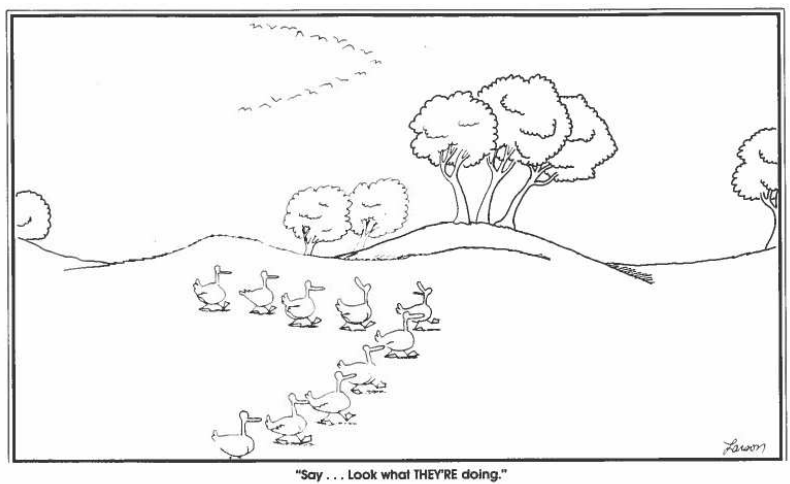}

    \caption{This irresistable {\it Far Side} cartoon by the
    magnificent Gary Larson does not quite express the premise of this
    BoF. To illustrate that the Users, Developers and Managers of scientific software are holding
    each other back by the way they are organized, the formation of
    the groundbound waddlers should of course be different from the lofty flyers. But we trust that you get the point anyway.}

\label{fig:V_ducks}
\end{figure}

\begin{abstract}

    This BoF is a continuation of the ADASS FADS tradition of yore,
    which aims to stimulate discussion (or at least awareness) about
    the non-technical aspects of our trade. This year, as we
    expected, it proved to be difficult to have a real discussion by
    means of Zoom. But the Metis polls and the DisCord log show that
    we have nevertheless been succesful in getting people to (begin
    to) think about the way Users, Developers and Managers work
    together, inside and between our MultiVerse of Bubbles. Next year we will try to focus this discussion by means of a Proxy demonstrator. We feel
    that there is a world to be won.

\end{abstract}

\section{Introduction}

Once upon a time, for a number of years, the final afternoon of the
ADASS conference was devoted to a wide-ranging discussion about the
{\it Future of Astronomical Dataprocessing Systems} (FADS). This was
very popular because software developers, in their quiet way, are
profound thinkers, and highly opiniated. The format was simple: a
premise was proposed, and after that the only problem was to gently
persuade orators to give someone else a chance.

It is not clear why the FADS tradition was discontinued, but given the
importance of ADASS, that question itself would be worthy of a FADS
discussion. In any case, we have quietly resurrected the tradition, in
the somewhat sub-optimal form of a yearly BoF session. Last year the
theme was {\it Escaping from the Herd of White Elephants}, and this year
we talk about {\it Interoperability}.

  The central idea of FADS is to get people to briefly rise above the
  fascinating {\it technical} challenges of our trade, and to consider
  the way we work together. More specifically, to ponder the pros and
  cons of making things as easy as possible for everyone. 
  Drastically lowering the threshold for many more brains and eyeballs to enter the fray {\it in a disciplined manner}
  should greatly accelerate the evolution of our field\footnote{Such inclusiveness
  was the engine behind the greatest leaps in human progress, e.g. 500
  BC, 1500 AD, 1863, and possibly 1994}.

It is difficult enough to keep a room full of software developers on
the subject\footnote{Just another example of trying to herd cats}, but
this year presented some extra challenges.  Although Zoom may be
satisfactory for giving an online presentation, and to answer a few
questions, it is definitely unsatisfying for a group discussion.
Nevertheless, the Metis polls and the DisCord log show that it is not
hard to get people to think critically about the way Users, Developers
and Managers of scientific software work together, in our MultiVerse
of Bubbles.

\section{This year's Premise}

\begin{wrapfigure}{l}{0.15\textwidth}
    \centering
    \includegraphics[width=0.15\textwidth]{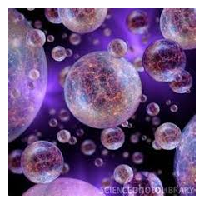}
\end{wrapfigure}

The short version of our premise is that we are held back by a {\it
Multiverse of Bubbles}. A Bubble may be generated by the love for an instrument like
a major telescope, a software environment, a user interface or
whatever.  Any attempt to get Users, Developers and Managers to work
better inside a Bubble, makes it harder for them to work with other Bubbles. As
a result, the field evolves more slowly than we can afford.

Of course the simple realization that Bubbles exist, and are counter-productive, should already have an effect. So that was our first
goal. But for a useful discussion about what might be done, a little
more focus is required. For this we encourage the reader of this summary to refer to
the slides that were used at the BoF, which should be available by
now (\url{https://adass2020.es/static/ftp/B11-143/B11-143_v3.pdf}).


\section{The polls of the BoF Participants}

\begin{wrapfigure}{l}{0.15\textwidth}
    \centering
    \includegraphics[width=0.15\textwidth]{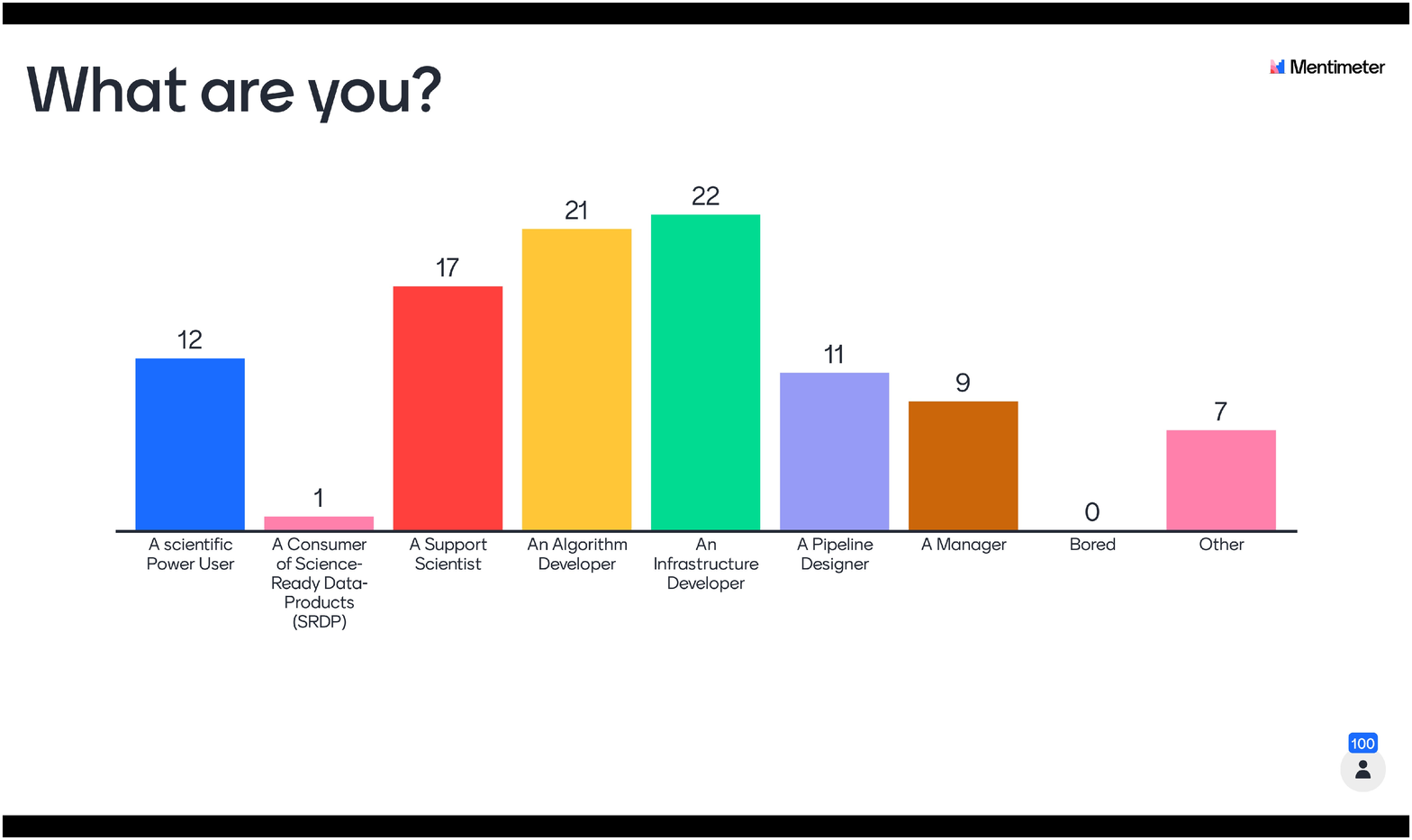}
\end{wrapfigure}

Since the idea was to get participants to introspect, we decided to
experiment with the polling feature of Zoom.  Unfortunately, this
feature was disabled by ADASS, but we managed to use Metis instead. We
had three polls, and only one answer could be given for each item.
\\ 

\begin{itemize}

  \item {\bf What are you? (100):} A scientific Power User (12), a
  Consumer of Science-Ready Data Products (1), a Support scientist
  (17), an Algorithm developer (21), an Infrastructure developer (22),
  A Pipeline designer (11), a Manager (9), bored (0!), Other (7).

  \item {\bf What kind of User are you? (104):} A Power User (33), a
  Consumer of SRDPs (4), a Victim (7), a Support scientist (27), a
  Tester (13), Other (20).

  \item {\bf Why I stick to my bubble (83):} I get paid (42), I hate
  learning another User Interface (3), it is the bubble of my PhD
  supervisor (1), it is good enough for my needs (12), Other (12), I
  do not understand the question (13).

\end{itemize}

At the very least these results emphasize that, in looking for ways to
work better together, we should be aware that there are different
kinds of Users and Developers, who may face different problems. For instance, consider the plight of the long-suffering and under-valued Support Scientist.

\section{The attempt at an online Discussion}

It is not easy to have a free-ranging discussion by means of Zoom.
Apart from some lapses in our own preparation\footnote{Our moderator
could not speak with his headphones plugged in, because it did not
interface with the Linux of his laptop. But without headphones, he
could not hear the speaker very well, so it was difficult to lead the
discussion.}, the inevitable extra formality gets in the way. So we
disconsolately plowed through the copious input slides, in the hope
that these would at least trigger food for thought in individuals. But
the moderator did not realize that, while only a few participants
spoke up on Zoom, many were diligently typing away on DisCord. This
gave us a valuable written record of 179 comments. Here are some of
the issues that were raised most frequently:

\begin{itemize}

  \item Recognition, also for infrastructure software, testing and support

  \item Other incentives, like career tracks and access to funding  

  \item The various kinds of Bubbles, and how to transcend them

  \item Getting time to explore, rather than just produce (like Google
    employees), e.g. for learning about other Bubbles

  \item The nature of Proxies, and whether they already exist

  \item And a lot of technical issues, which were of course outside
  the scope of this BoF

\end{itemize}

The reader is encouraged to work through these DisCord comments, and
perhaps contact those who made them. In any case, the comments are already
helping us to sharpen the Proxy story (see below), in such a way that they address the expressed concerns more explicitly, and more up-front.

\section{Next year's subject: Proxies}

The second half of the BoF input slides sketch a system based on
Proxies. It addresses many of the issues that have come up in this
BoF. Apart from being a real proposal, it may also be used as a means
to focus next year's discussion a little more.

The Proxy system proposes that each {\it compute resource} (software,
data, computer etc) may also be accessed by means of specific Proxy objects, which know all the gory details of its underlying
target module, and takes care of everything that a User should not
have to deal with. Including remote processing and containers, of course. In addition, Proxies relieve the Developer of much
(Bubble-specific!) infrastructure work, and make their stuff widely
available. Since all Proxies share a uniform interface, they can be
combined in ``pipeline'' Proxies that can do a lot more than current
pipelines. A single GUI becomes {\bf the one Bubble that rules them all}, because it offers one-click access to a huge number of
Proxy objects that use  existing compute resources in
arbitrary ways.

The many small Proxy definition files (Python code, probably) are kept in a central repository
that is looked after by a group whose role resembles that of the
Python core team. It provides invaluable services like making sure
that regular testing, and maintenance is done etc. The Prime Directive
states that {\it no Proxy that is accepted by the repository can ever
be changed}. Instead, they may used to generate new Proxies by cloning
(e.g. with different input arguments) or derivation (with  modified code).

Much more may be said about Proxies, and hopefully will, whether they get implemented or
not. But in the meantime we cannot resist to emphasize one feature here that should strongly resonate with the DisCord comments: {\bf Proxies
can calculate the relative contributions of the various developers of
a Proxy, including its inheritance tree! This may be used to control a
flow of incentives (e.g. money, or praise), weighed by the number of times the
Proxy is used, i.e. downloaded from the repository.}

Finally, innovation is often obstructed because it is not in the
interest of those that must make it happen (we trust that you can
think of your own examples). So revolutions tend to be caused by
solutions that simply bypass such obstacles. Proxies are such a
solution: A small group can quickly implement a GUI, and generate a few sample
Proxies that use popular existing software to do some interesting
things. And then let the system prove itself. 

More next year. We are
convinced that there is a world to be won.

 

\end{document}